\documentclass[twocolumn]{aa}
\usepackage{graphicx}
\usepackage{txfonts}
\usepackage[authoryear]{natbib}
\bibpunct{(}{)}{;}{a}{}{,}
\newcommand{\sax}  {SAX\,J2239.3+6116}
\newcommand{\ha}  {H$\alpha$}
\newcommand{\ew}  {EW(H$\alpha$)}

\def\simless{\mathbin{\lower 3pt\hbox
     {$\rlap{\raise 5pt\hbox{$\char'074$}}\mathchar"7218$}}}   
\def\simmore{\mathbin{\lower 3pt\hbox
     {$\rlap{\raise 5pt\hbox{$\char'076$}}\mathchar"7218$}}}   

\begin{document}

   \title{The optical counterpart to the Be/X-ray binary SAX\,J2239.3+6116}

   \subtitle{}
  \author{
        P. Reig\inst{1,2}
        \and
        P. Blay\inst{3,4}
        \and
        D. Blinov\inst{2,5}
           }

\authorrunning{Reig et al.}
\titlerunning{The optical counterpart to \sax}

   \offprints{pau@physics.uoc.gr}

   \institute{IESL, Foundation for Reseach and Technology-Hellas, 71110, 
                Heraklion, Greece. 
         \and Physics Department, University of Crete, 71003, 
                Heraklion, Greece.
                \email{pau@physics.uoc.gr}
         \and Instituto de Astrof\'{\i}sica de Canarias, Tenerife, Spain.
                \email{pblay@iac.es}
         \and Nordic Optical Telescope, La Palma, Spain.
         \and Astronomical Institute, St. Petersburg State University,
         Universitetsky pr. 28, Petrodvoretz, 198504 St. Petersburg,
         Russia.
        }

   \date{Received ; accepted}

\abstract
{Be/X-ray binaries represent the main group of high-mass X-ray binaries.
The determination of the astrophysical parameters of the counterparts of
these high-energy sources  is important for the study of X-ray binary
populations in our Galaxy. X-ray observations suggest that
SAX\,J2239.3+6116 is a Be/X-ray binary. However, little is known about the
astrophysical parameters of its massive companion.
}
{The main goal of this work is to perform a detailed study of the optical
variability of the Be/X-ray binary SAX\,J2239.3+6116. }
{We obtained multi-colour $BVRI$ photometry and polarimetry and  4000-7000 \AA\
spectroscopy.  The 4000--5000 \AA\ spectra allowed us to
determine the spectral type and projected rotational velocity of the optical companion; 
the 6000-7000 \AA\ spectra, together with the photometric magnitudes, were used to
derive the colour excess $E(B-V)$, estimate the distance, and to study
the variability of the H$\alpha$ line.  }
{The optical counterpart to SAX J2239.3+6116 is a $V=14.8$ B0Ve star
located at a distance of $\sim$4.9 kpc. The interstellar reddening in the
direction of the source is $E(B-V)=1.70\pm0.03$ mag. The monitoring of the H$\alpha$
line reveals a slow long-term decline of its equivalent width since 2001.
The line profile is characterized by a stable double-peak profile with no
indication of large-scale distortions. We measured intrinsic
optical polarization for the first time. Although somewhat higher than
predicted by the models, the optical polarization is consistent with
electron scattering in the circumstellar disk. } 
{We attribute the long-term decrease in the intensity of the H$\alpha$ line to 
the dissipation of the circumstellar disk of the Be star. The longer variability 
timescales observed in SAX\,J2239.3+6116 compared to other Be/X-ray binaries 
may be explained by the wide orbit of the system.}

\keywords{stars: individual: \sax,
 -- X-rays: binaries -- stars: neutron -- stars: binaries close --stars: 
 emission line, Be
               }

   \maketitle

\begin{table*}
\caption{\ha\ line parameters of the optical counterpart to \sax\ : we give the
equivalent width, peak separation, and $V/R$ ratio. }
\label{spec}
\begin{center}
\begin{tabular}{lccccc}
\hline  \hline
Date            &JD             &Telescope      &\ew\     &$\Delta_{\rm p}$     &$\log(V/R)$   \\
                &               &               &(\AA)    &(km s$^{-1}$)   &         \\     
\hline\hline
31-05-2001      &2452061.558    &SKT    &$-8.34\pm0.22$     &162$\pm$10      &$-0.032\pm0.030$\\
03-06-2001      &2452064.479    &SKT    &$-8.80\pm0.22$      &150$\pm$10      &$-0.008\pm0.030$\\
13-09-2001      &2452166.417    &SKT    &$-10.52\pm1.25$    &202$\pm$10      &$-0.004\pm0.012$\\
08-10-2001      &2452191.367    &SKT    &$-10.02\pm0.24$    &175$\pm$10      &$-0.006\pm0.030$\\
18-07-2002      &2452474.481    &SKT    &$-7.98\pm0.39$     &--              &-- \\
08-10-2003      &2452921.323    &SKT    &$-10.17\pm0.28$    &--              &-- \\
24-06-2004      &2453181.465    &SKT    &$-10.11\pm0.33$    &159$\pm$10      &$-0.058\pm0.003$\\
24-06-2005      &2453546.433    &SKT    &$-8.74\pm0.31$     &222$\pm$5       &$-0.059\pm0.014$\\
13-07-2005      &2453565.522    &SKT    &$-8.96\pm0.23$     &241$\pm$4       &$-0.059\pm0.010$\\
17-08-2005      &2453600.387    &SKT    &$-8.47\pm0.39$     &240$\pm$5       &$-0.020\pm0.011$\\
26-10-2005      &2453670.388    &SKT    &$-7.97\pm0.22$     &171$\pm$10      &$-0.064\pm0.030$\\
06-09-2007      &2454350.508    &SKT    &$-5.97\pm0.18$     &258$\pm$8       &$-0.046\pm0.018$\\
25-06-2008      &2454643.486    &SKT    &$-6.74\pm0.20$     &238$\pm$10      &$-0.021\pm0.030$\\
03-09-2008      &2454713.324    &SKT    &$-7.53\pm0.19$     &252$\pm$7       &$-0.009\pm0.015$\\
12-08-2009      &2455056.477    &SKT    &$-4.98\pm0.31$     &241$\pm$4       &$-0.048\pm0.010$\\
03-08-2010      &2455412.382    &SKT    &$-5.58\pm0.21$     &279$\pm$7       &$ 0.019\pm0.016$\\
29-08-2010      &2455438.417    &SKT    &$-5.66\pm0.14$     &255$\pm$9       &$-0.047\pm0.021$\\
30-09-2010      &2455470.437    &SKT    &$-5.58\pm0.13$     &259$\pm$10      &$-0.007\pm0.023$\\
21-08-2011      &2455795.414    &SKT    &$-5.97\pm0.13$     &243$\pm$6       &$-0.048\pm0.013$\\
24-08-2012      &2456164.453    &SKT    &$-4.09\pm0.16$     &292$\pm$7       &$ 0.025\pm0.017$\\
06-09-2012      &2456177.505    &SKT    &$-4.32\pm0.16$     &289$\pm$9       &$-0.006\pm0.022$\\
14-09-2012      &2456185.419    &SKT    &$-3.66\pm0.18$     &279$\pm$4       &$ 0.009\pm0.010$\\
19-10-2012      &2456220.352    &SKT    &$-3.91\pm0.18$     &286$\pm$7       &$-0.022\pm0.012$\\
30-07-2013      &2456504.476    &SKT    &$-5.31\pm0.21$     &245$\pm$5       &$ 0.033\pm0.007$\\
31-08-2013      &2456536.413    &SKT    &$-6.13\pm0.29$     &249$\pm$4       &$-0.003\pm0.006$\\
19-10-2013      &2456585.298    &SKT    &$-6.28\pm0.24$     &252$\pm$3       &$-0.035\pm0.006$\\
07-08-2014      &2456877.512    &SKT    &$-1.91\pm0.11$     &345$\pm$5       &$ 0.089\pm0.019$\\
15-09-2014      &2456915.609    &NOT    &$-2.11\pm0.21$     &344$\pm$1       &$ 0.036\pm0.014$\\
12-10-2014      &2456943.391    &SKT    &$-2.15\pm0.27$     &310$\pm$5       &$ 0.059\pm0.015$\\
23-06-2015      &2457197.500    &SKT    &$-1.01\pm0.21$     &343$\pm$6       &$ 0.093\pm0.028$\\
07-07-2015      &2457211.500    &SKT    &$-0.78\pm0.19$     &349$\pm$7       &$ 0.004\pm0.032$\\
04-08-2015      &2457239.500    &WHT    &$-1.30\pm0.41$     &379$\pm$2       &$ 0.010\pm0.009$\\
05-10-2015      &2457301.500    &SKT    &$-2.34\pm0.33$     &347$\pm$8       &$ 0.055\pm0.019$\\
06-10-2015      &2457302.500    &SKT    &$-2.16\pm0.25$     &317$\pm$4       &$ 0.019\pm0.015$\\
06-11-2015      &2457333.501    &NOT    &$-1.84\pm0.90$     &312$\pm$2       &$ 0.097\pm0.008$\\
26-05-2016      &2457534.635    &NOT    &$-2.50\pm0.90$     &303$\pm$10      &$-0.055\pm0.008$\\
\hline\hline
\end{tabular}
\end{center}
\end{table*}

\begin{table*}
\caption{Multi-colour photometry and polarimetry of the optical counterpart to \sax.}
\label{phot}
\begin{center}
\begin{tabular}{cccccc}
\noalign{\smallskip}    \hline\noalign{\smallskip}
\multicolumn{6}{c}{Photometry (mag.)}\\
\noalign{\smallskip}    \hline\noalign{\smallskip}
Date &  JD (2,400,000+)    &   $B$  &   $V$   &   $R$  & $I$   \\
\noalign{\smallskip}    \hline\noalign{\smallskip}
16-07-2007      &54298.518      &16.25$\pm$0.02 &14.75$\pm$0.03 &13.85$\pm$0.02 &12.86$\pm$0.03   \\        
30-06-2009      &55013.513      &16.26$\pm$0.02 &14.79$\pm$0.01 &13.91$\pm$0.01 &12.97$\pm$0.02   \\        
26-08-2011      &55800.488      &16.30$\pm$0.01 &14.83$\pm$0.01 &13.95$\pm$0.01 &13.02$\pm$0.02   \\        
09-09-2011      &55814.349      &16.28$\pm$0.03 &14.87$\pm$0.03 &14.00$\pm$0.02 &13.05$\pm$0.03   \\       
29-08-2013      &56534.500      &16.29$\pm$0.02 &14.84$\pm$0.02 &13.90$\pm$0.02 &12.91$\pm$0.03   \\        
20-08-2014      &56890.479      &16.23$\pm$0.02 &14.81$\pm$0.02 &13.91$\pm$0.02 &12.96$\pm$0.03   \\
22-07-2015      &57226.536      &16.30$\pm$0.02 &14.84$\pm$0.01 &13.97$\pm$0.01 &12.97$\pm$0.01   \\
18-11-2015      &57345.326      &16.26$\pm$0.02 &14.82$\pm$0.01 &13.94$\pm$0.02 &13.00$\pm$0.02   \\
06-06-2016      &57546.580      &16.26$\pm$0.02 &14.82$\pm$0.02 &13.94$\pm$0.02 &12.99$\pm$0.03   \\
\noalign{\smallskip}    \hline\noalign{\smallskip}
\multicolumn{6}{c}{Polarimetry (\%)}\\
\noalign{\smallskip}    \hline\noalign{\smallskip}
10-07-2015      &57214.541      &7.8$\pm$1.9    &7.3$\pm$0.6    &6.8$\pm$0.3    &6.1$\pm$0.3    \\
15-06-2016      &57555.467      &7.5$\pm$0.8    &7.4$\pm$0.2    &6.7$\pm$0.3    &6.3$\pm$0.2    \\      
\noalign{\smallskip}    \hline
\end{tabular}
\end{center}
\end{table*}
   
\begin{table*}
\caption{Polarization degree, polarization angle, and Stokes parameters of the optical counterpart to \sax\ 
in the $R$ band.}
\label{pol}
\begin{center}
\begin{tabular}{lccccc}
\noalign{\smallskip}    \hline\noalign{\smallskip}
Date &  JD (2,400,000+)         &PD(\%)         &PA ($^\circ$)          &$q$                    &$u$     \\
\noalign{\smallskip}    \hline\noalign{\smallskip}
06-10-2013   &56572.458      &7.5$\pm$0.4     &63.3$\pm$1.6   &-0.045$\pm$0.004    &0.060$\pm$0.004   \\ 
08-10-2013   &56574.345      &6.3$\pm$0.4     &63.2$\pm$1.7   &-0.037$\pm$0.004    &0.051$\pm$0.004   \\
20-10-2013   &56586.360      &6.8$\pm$0.3     &63.3$\pm$1.1   &-0.040$\pm$0.003    &0.054$\pm$0.003   \\
28-10-2013   &56594.444      &6.8$\pm$0.2     &61.7$\pm$1.0   &-0.037$\pm$0.003    &0.057$\pm$0.003   \\
10-11-2013   &56607.349      &6.6$\pm$0.3     &63.1$\pm$1.4   &-0.039$\pm$0.003    &0.053$\pm$0.003   \\
23-05-2014   &56801.571      &6.5$\pm$0.4     &61.1$\pm$1.6   &-0.034$\pm$0.004    &0.055$\pm$0.004   \\
14-07-2014   &56853.600      &6.8$\pm$0.4     &61.9$\pm$1.5   &-0.038$\pm$0.004    &0.057$\pm$0.004   \\
02-08-2014   &56872.535      &6.7$\pm$0.4     &64.1$\pm$1.8   &-0.042$\pm$0.004    &0.053$\pm$0.004   \\
14-08-2014   &56884.584      &6.9$\pm$0.4     &62.9$\pm$1.9   &-0.040$\pm$0.005    &0.056$\pm$0.005   \\
01-09-2014   &56902.487      &6.8$\pm$0.4     &64.5$\pm$1.9   &-0.043$\pm$0.004    &0.053$\pm$0.004   \\
10-07-2015   &57214.541      &6.8$\pm$0.3     &63.2$\pm$1.3   &-0.040$\pm$0.003    &0.054$\pm$0.003   \\
05-08-2015   &57240.569      &6.8$\pm$0.4     &64.8$\pm$1.5   &-0.044$\pm$0.004    &0.053$\pm$0.004   \\
19-11-2015   &57346.284      &6.9$\pm$0.3     &63.6$\pm$1.3   &-0.042$\pm$0.003    &0.055$\pm$0.003   \\
14-06-2016   &57554.580      &6.7$\pm$0.2     &64.3$\pm$1.0   &-0.042$\pm$0.002    &0.052$\pm$0.002   \\
15-06-2016   &57555.451      &6.6$\pm$0.3     &62.9$\pm$1.1   &-0.038$\pm$0.003    &0.053$\pm$0.003   \\
\noalign{\smallskip}   \hline       
\end{tabular}
\end{center}
\end{table*}

\section{Introduction}

\sax\ was discovered by the {\em BeppoSAX} wide-field camera as a transient
X-ray source during  observations of the supernova remnant Cas A. It  was
first detected on 4 March 1997 and then again on 8 May 1999
\citep{intzand00}. The peak flux then was $3.3 \times 10^{-10}$ erg
s$^{-1}$ cm$^{-2}$ in the energy range 2--10 keV and $1.0 \times 10^{-9}$
erg s$^{-1}$ cm$^{-2}$ in the energy range 2--26 keV. The X-ray spectral
continuum was satisfactorily fitted with a single absorbed power law with
$N_H=1\times 10^{22}$ cm$^{-2}$ and $\Gamma=1.1\pm0.1$. The source was not
detected on 13--15 December 1999 with an upper limit on the 2--10 keV X-ray
 flux of $6 \times 10^{-11}$ erg s$^{-1}$ cm$^{-2}$ \citep{intzand00}.
A search for detections with other X-ray instruments resulted in a
detection with {\em BeppoSAX}/MECS on 24--25 November 1998 with a flux of 
$5 \times 10^{-13}$ erg s$^{-1}$ cm$^{-2}$ (2--10 keV) and another detection with
{\em CGRO}/BATSE in March 1997 with a flux $1.4 \times 10^{-9}$ erg
s$^{-1}$ cm$^{-2}$ (20--100 keV). The {\em RXTE}/ASM light curve  showed
increases in the X-ray intensity at regular interval times of $262\pm5$
days \citep{intzand00}.   If this periodicity is interpreted as the orbital
period of the system, then \sax\ has the longest orbital period of all the
known BeXB in our Galaxy.  Subsequent {\em RXTE}/PCA and {\em
BeppoSAX}/MECS-LECS observations during a predicted outburst in July 2001
revealed X-ray pulsations with a pulse period of $1247.2\pm0.7$ s
\citep{intzand01}. The X-ray flux during the observations that detected
pulsations was $4 \times 10^{-12}$ erg s$^{-1}$ cm$^{-2}$ (2--10 keV). 

Optical observations were carried out on 2--3 December 1999 with the 2.1 m
telescope of the Kitt Peak National Observatory. A B-type star showing
\ha\ in emission with an equivalent width of --6.7 \AA\ was discovered
$0.3\arcmin$ away from the best-fit X-ray position. This V=15.1 mag star
was proposed to be the optical counterpart to \sax\ \citep{intzand00}.

One more optical campaign was reported by \citet{riquelme12} on 2--5 July
2001 with the following photometric magnitudes: $U=16.19$, $B=16.06$,
$V=14.55$, $R=13.60$, and $I=12.74,$ and an \ha\ equivalent width of --11.0
\AA.

In this work, we performed the first detailed study of the optical
variability of \sax. We present photometric observations covering the
period 2007-2016, spectra in the region of the \ha\ line from 2001-2016, and
for the first time polarimetric data from 2013-2016. Although photometric
and spectroscopic data prior to 2015 have been presented in the context of
a global study of the long-term optical variability of BeXBs by \citet{reig15}
and \citet{reig16}, respectively, we include them here for the sake of
completeness. Our polarimetric
observations are the first dedicated observations of the source using this
technique.

\section{Observations}

\subsection{Spectroscopy}

Optical spectroscopic observations were obtained from the 1.3 m telescope
of the Skinakas observatory (SKT) in Crete (Greece) and from the 2.5 m
Nordic Optical Telescope (NOT) in El Roque de los Muchachos observatory (La
Palma, Spain).  \sax\ was also observed in service time with the 4.2 m 
William Herschel Telescope (WHT). 

The 1.3\,m telescope of the Skinakas Observatory was equipped with a
2000$\times$800 ISA SITe CCD and a 1302 lines~mm$^{-1}$ grating, giving a
nominal dispersion of $\sim$1 \AA/pixel.  The NOT was equipped with the
Andalucia Faint Object Spectrograph and Camera (ALFOSC), an  EEV42-40, 2Kx2K
chip, and Grism\#16 (1000 lines~mm$^{-1}$; 0.86 \AA/pixel) for the blue
spectrum (3460--5220 \AA) on 16 July 2015 and Grism\#17 (2400
lines~mm$^{-1}$; 0.26 \AA/pixel) for the red spectra (6330-6870 \AA) on 15
September 2014 and 6 November 2015. The NOT red spectrum on 26 May 2016 was
made with Grism\#20 (484 lines~mm$^{-1}$, 5650-10150 \AA, 2.2 \AA/pixel).
The WHT was used with the ISIS spectrograph and an EEV12 CCD chip. The
blue arm covers the range 3927--4573 \AA\ with the R1200B gratings and
gives a dispersion of 0.23 \AA/pixel, while the red arms covers the range
6491-7109 \AA\ and gives a dispersion of 0.26 \AA/pixel.

After correcting the images for bias and flatfield, spectra were extracted
from the object and the nearby sky. The sky spectrum was subtracted from
that of the object and the resulting spectrum was wavelength-calibrated.
Spectra of comparison lamps were taken before each exposure to
account for small variations in the wavelength calibration during the
night. 

To ensure a homogeneous processing of the spectra, they were normalized
with respect to the local continuum, which  was rectified to unity by
employing a spline fit. The definition of the continuum level is crucial
because it represents the main source of uncertainty in the determination
of the spectral parameters.  

To extract the \ha\ line parameters, we fitted the profiles with two
Lorentzian functions. Lorentzian profiles behaved better than Gaussian
profiles owing to the extended wings and the small peak distance in spectra
with large values of \ew. Twelve different selections of the continuum were
used. For each one, we obtained the line centres, full width at half
maximum, and the intensity of the two peaks above the local continuum. The final
value and error of these parameters were computed as the average and
standard deviation of those 12 measurements. The equivalent width, peak
separation, and V/R ratio of the \ha\ line are given in Table~\ref{spec}.

\subsection{Photometry}

The photometric observations were made with the 1.3 m telescope of the
Skinakas Observatory. \sax\ was observed through the Johnson-Cousins $B$,
$V$, $R,$ and $I$ filters. For the photometric observations, the telescope
was equipped with a  2048$\times$2048 ANDOR CCD with a 13.5 $\mu$m pixel
size. In this configuration, the  plate scale is 0.28 $\arcsec$/pixel,
hence providing a field of view of $9.5 \times 9.5$ arcmin$^2$. Standard
stars from the Landolt list \citep{landolt09} were used for the
transformation equations.  Reduction of the data was carried out in the
standard way using the IRAF tools for aperture photometry. After the
standardization process, we finally assigned an error to the calibrated
magnitudes of the target given by the rms of the residuals between the
catalogued and calculated magnitudes of the standard stars. The photometric
magnitudes are given in Table~\ref{phot}. 

\subsection{Polarimetry}

Polarimetric observations in the $R$ band were made with the RoboPol
photopolarimeter attached to the focus of the 1.3 m telescope of the
Skinakas Observatory (Table~\ref{pol}).  In addition, we obtained 
multi-colour polarimetric data on the nights 10 July 2015 and 15 June  2016
(see Table~\ref{phot}).  In the polarimetry configuration a plate scale of 
0.43 $\arcsec$/pixel is achieved with a 2048$\times$2048 ANDOR CCD with a
13.5 $\mu$m pixel size.  RoboPol is an imaging photopolarimeter that
measures simultaneously the Stokes parameters of linear polarization  of
all sources in the field of view \citep{king14}.  Robopol splits the
incident light into two beams, each half incident on a half-wave retarder
followed by a Wollaston prism.  The fast axis of the half-wave retarder in
front of the first prism is rotated by $67.5^{\circ}$ with respect to the
other retarder. Every point in the sky is thereby projected to four points
on the CCD.  In each spot the photon counts, measured using aperture photometry, are used to calculate the $U$ and $Q$ parameters of linear
polarization.  A mask is placed in
the telescope focal plane  to optimize the instrument sensitivity. The absence of moving parts allows RoboPol to
compute all the Stokes parameters of linear polarization in one shot.

\begin{figure}
\resizebox{\hsize}{!}{\includegraphics{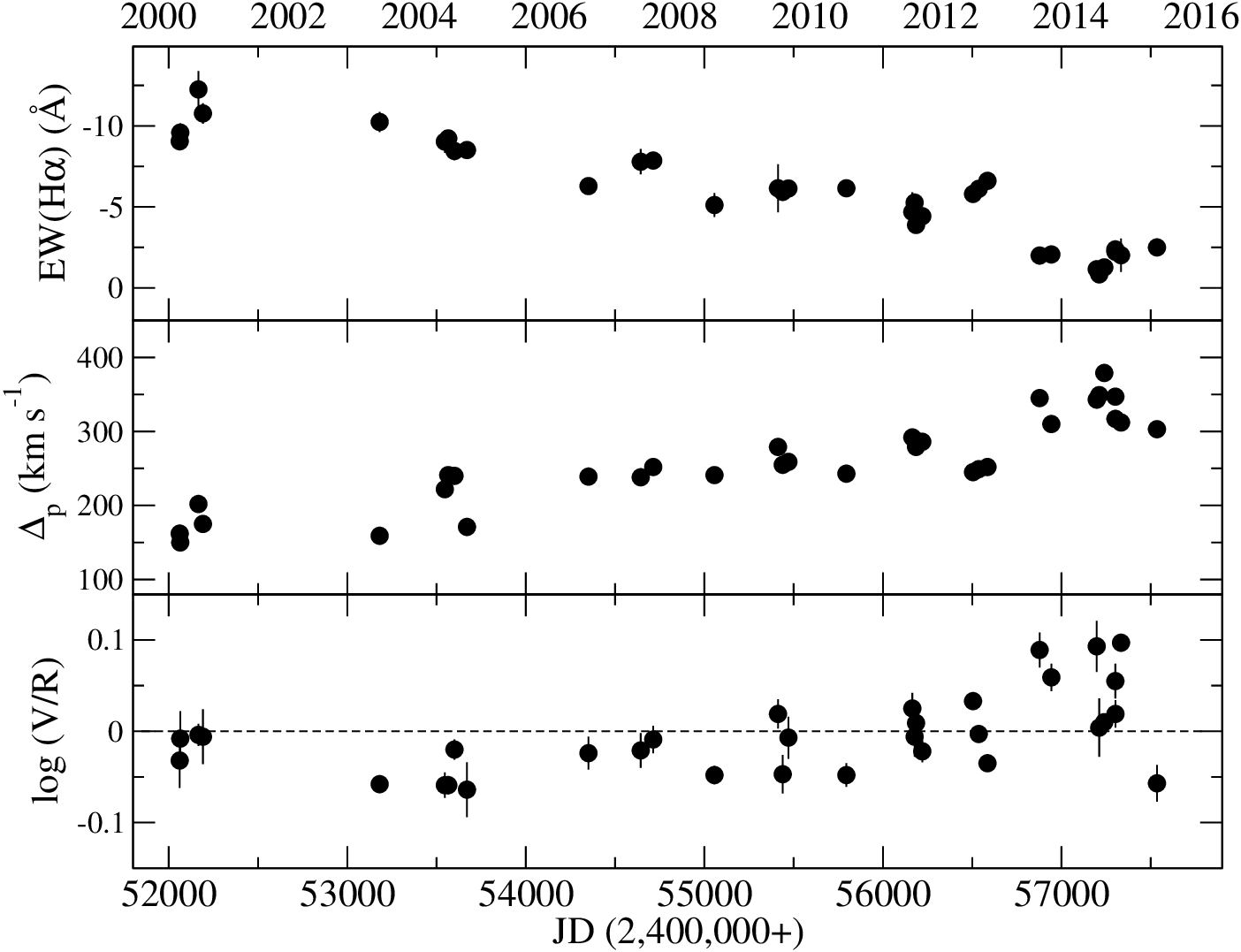}} 
\caption[]{Evolution of the \ha\ equivalent width, peak separation, and V/R
ratio.  }
\label{hapar}
\end{figure}
\begin{figure}
\resizebox{\hsize}{!}{\includegraphics{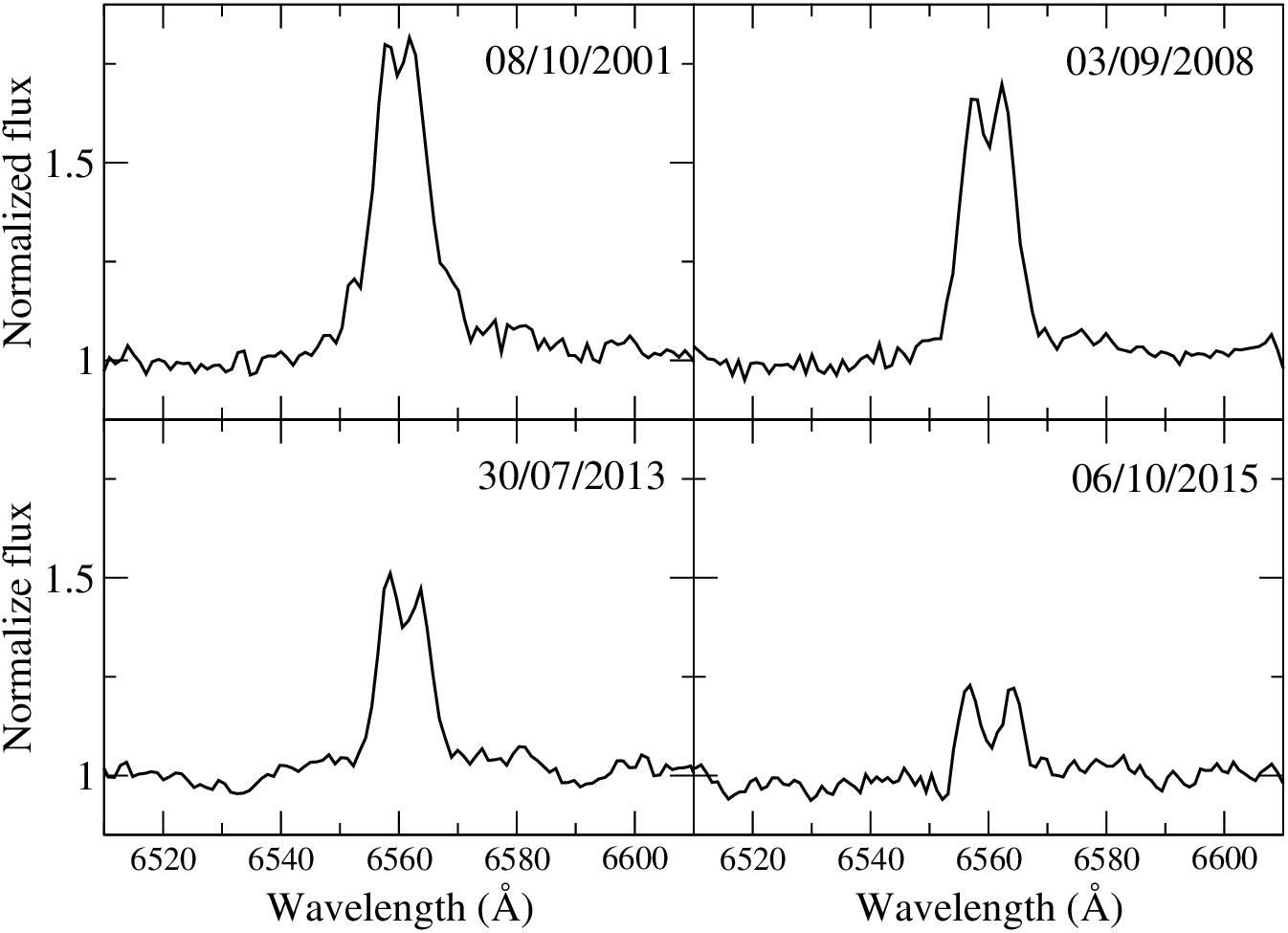}} 
\caption[]{Representative profiles of the \ha\ line. }
\label{haprof}
\end{figure}

\section{Results}

\subsection{The \ha\ line: Evolution of spectral parameters}
\label{haevol}

Table~\ref{spec} summarizes the results of the spectral analysis, where
the equivalent width of the \ha\ line (\ew), peak separation ($\Delta_{\rm p}$), and
V/R ratio ($\log(V/R)$) are given. The peak separation is simply the
difference between the central wavelength of the red minus the blue peak in
velocity units ($\Delta \lambda/\lambda \times c$), where $c$ is the speed
of light.   The errors are estimated by propagating the uncertainty in the
determination of the central wavelength of the Lorentzian profile used to fit
the \ha\ profile. The $V/R$ ratio is defined as the ratio of the relative
intensity at the blue and red emission peak maxima (after subtracting the
underlying continuum). For plotting purposes we used the log of this ratio,
$\log(V/R)$. Thus negative values indicate a red-dominated peak ($V < R$),
positive values indicate a blue-dominated profile ($V > R$), and values close to
zero correspond to equal intensity peaks ($V\approx R$).

The \ew\ of \sax\ shows a long-term decreasing trend since 2001. The \ew\ 
changed from $\sim$--12 \AA\ in September 2007 to $\sim$--1 \AA\ in July
2015 (Fig.~\ref{hapar}, top panel). The decreasing trend is not smooth but
displays a saw-tooth profile, i.e. sometimes \ew\ increases to subsequently
drop below the initial trend. The changes in the shape of the line are also
significant. Although  $V \approx R$ is the dominant profile, there is a
long-term trend of the $V/R$ ratio turning from a slightly red-dominated
profile ($V/R \simless 1$) into a slightly blue-dominated profile ($V/R
\simmore 1$) as the intensity of the line decreases (Fig.~\ref{hapar},
bottom panel). The peak separation also changes with time. It is smaller
when \ew\ is larger (Fig.~\ref{hapar}, middle panel).

Fig.~\ref{haprof} figure shows \ha\ profiles at four different epochs. In
the spectra with \ew $\simmore$ 10 \AA, the peaks are so close together
that the fits with two Lorentzian profiles do not always  provide sensible
values. In these cases we simply obtained $\Delta_{\rm p}$ by visual
inspection, finding the wavelength of the local maxima directly from the
data.

\begin{figure*}
\begin{center}
\includegraphics[width=16cm,height=10cm]{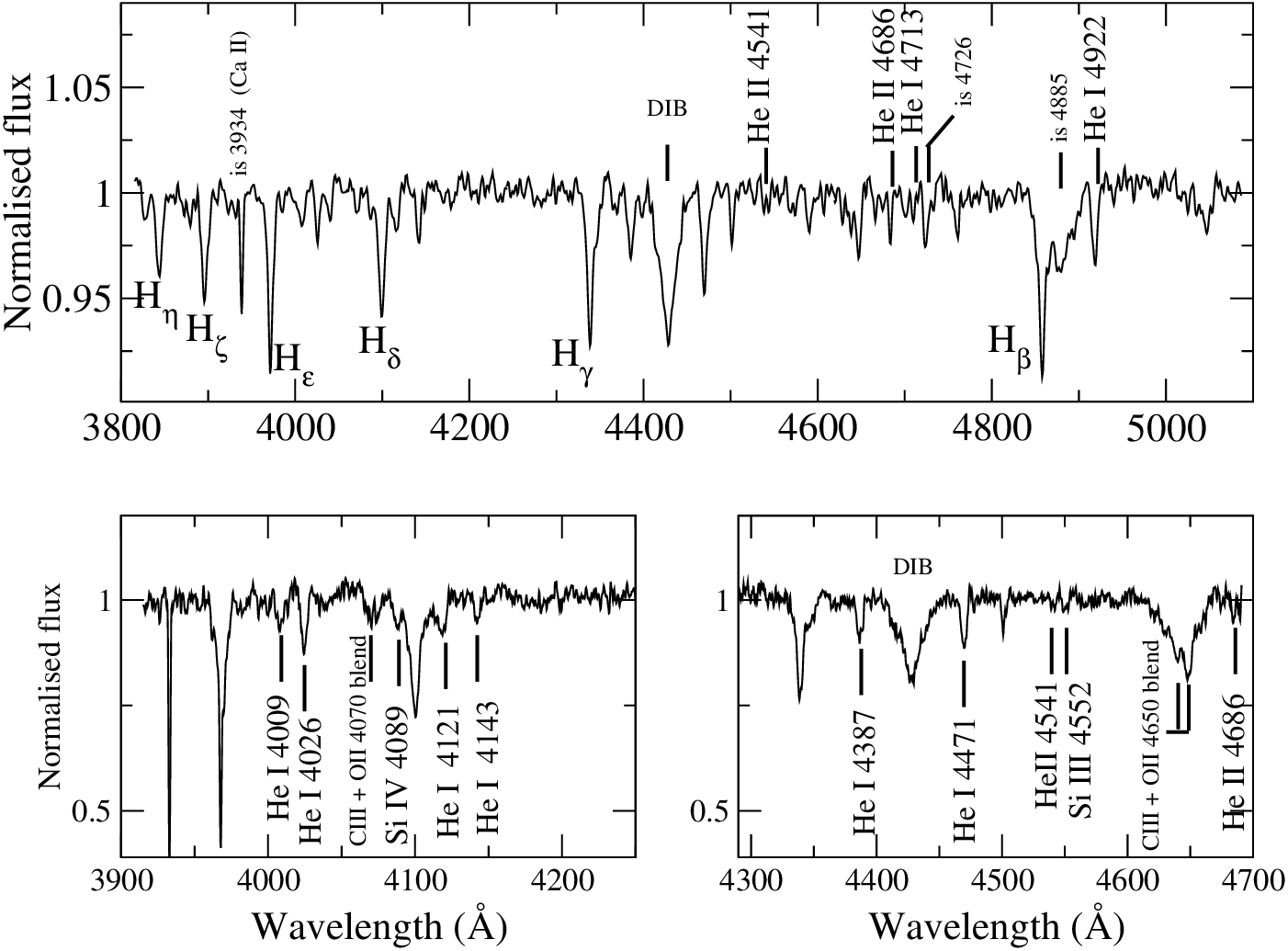} 
\caption[]{Identification of the spectral lines used for 
classification.  The upper panel shows the NOT spectrum smoothed with a
Gaussian filter of $\sigma=1 \AA$. The lower panels show the WHT spectrum 
smoothed with a Gaussian filter of $\sigma=0.3 \AA$.}
\label{speclass}
\end{center}
\end{figure*}

\subsection{Spectral classification}

The spectral type of \sax\ is not well constrained. The low resolution
spectra ($\Delta \lambda=5$ \AA) of the 1999 optical observations showed H
and He I lines and numerous diffuse interstellar bands (DIB), suggesting a
B0V--B2III star \citep{intzand00}. Our higher resolution spectra reveals
many more features. Figure~\ref{speclass} shows the NOT ($\Delta
\lambda=2.7$ \AA) and WHT ($\Delta \lambda=0.9$ \AA) spectra in the
traditional classification region. The spectral resolution was
estimated by measuring the FWHM of a calibration lamp line.  For
plotting purposes the spectra shown in Fig.~\ref{speclass} have been
smoothed with a Gaussian filter ($FWHM=2.354*\sigma$) with $\sigma=1 \AA$
and $\sigma=0.3 \AA$, respectively.

The blue spectrum of \sax\  is dominated by hydrogen and neutral helium
absorption lines, clearly indicating an early-type B star. The Balmer
series lines from H$\beta$ up to H$\eta$ at 3835 \AA\ are seen in
absorption. Very weak He II lines ($\lambda$4541, $\lambda$4686) are also
present, which implies a type later than O9. Mg II $\lambda$4481 is absent
or very weak, indicating a type earlier than B1.  The strong  C III + O II
blends at 4070 \AA\ and 4650 \AA\ also support a spectral type between B0
and B1. The similar intensity of Si IV $\lambda$4089 with respect to the
nearby He I lines at 4121 \AA\ and 4143 \AA\ and of Si III  $\lambda$4552
with respect to He II $\lambda$4541 favours a B0 main-sequence spectral
class. A later spectral type, for example B1, would have the ratio He I
$\lambda$4121/Si IV $\lambda$4089 $>>$ 1, whereas an earlier type star,
for example O9, would have Si III  $\lambda$4552/He II $\lambda$4541 $<<$ 1
\citep{walborn90}. Likewise, the strength of Si IV $\lambda$4089 increases
quickly with luminosity and becomes larger than He I $\lambda$4121 and
$\lambda$4143 for giants and supergiants stars. The absence of OII lines
also indicates a luminosity class V star.  We conclude that the most likely
spectral type of \sax\ is B0Ve.

\subsection{Reddening and distance}
\label{red}

To estimate the distance, the amount of interstellar extinction $A_{\rm
V}=R \times E(B-V)$ to the source has to be determined.  It is well known
that the circumstellar disk in Be stars introduces extra reddening that
should be taken into account to determine the colour excess $E(B-V)$ from
the interstellar medium. \citet{riquelme12} provided a method to correct
for the excess emission from the disk. In fact, \sax\ was one of the
sources studied by these authors. At the time of their study, \sax\
exhibited strong \ha\ emission with \ew=--11 \AA. They obtained
$E(B-V)=1.66\pm0.05$. However, as shown in Sect.~\ref{haevol}, the strength
of the \ha\ line \sax\ has been decreasing since 2001 (Fig.~\ref{hapar}).
The source appears to be approaching a disk-loss state. The smallest value
of \ew\ is $\sim0.8$\AA, which corresponds to a disk contribution of
$E^{\rm cs}(B-V)\simless 0.01$ mag \citep{riquelme12}, which lies within
our photometric errors. 

We estimated the reddening to the star in two different ways. The
refinement of the spectral type allows us to use this fundamental spectral
information in combination with the photometric observations.  The observed
colour of \sax\ is $(B-V)=1.44\pm0.03$ (Table~\ref{phot}), while the
expected one for a B0V star $(B-V)_0=-0.29$
\citep{johnson66,gutierrez-moreno79,wegner94}.
Thus we derive a colour excess of $E(B-V)=1.73\pm0.03$. 

The second method uses the strength of DIBs
in the spectrum  \citep{herbig75,herbig91,galazutdinov00,
puspitarini13,kos13}. In this method, a least-squares fit is performed
between the equivalent width of DIBs measured in the spectra of a large
number of stars and the colour excess $E(B-V)$ of those stars,
$EW(\lambda)=a * E(B-V)+ b$. We used the coefficients given by
\citet{herbig75} because the sample of stars in this work covers a wider
range in $E(B-V)$. Table~\ref{dib} gives the details of the calculation.
The reddening derived for each line was obtained by considering the average
over all the spectra for which the line could be measured, and $N$ is the
number of measurements. The weighted mean of the four colour excesses is
$E(B-V)=1.58\pm0.18$ and it is consistent with the photometric derived
value. 

Taken $E(B-V)=1.70\pm0.05$ and assuming the standard extinction law $R=3.1$
and an average absolute magnitude for a B0V star of $M_V=-3.88\pm0.3$
\citep{vacca96,wegner06}, the distance to \sax\ is estimated to be  4.9$\pm$0.8
kpc. Our value agrees with previously reported distances:
\citet{riquelme12} found 5.4$\pm$0.5 kpc and \citet{intzand00} found 4.4 kpc.

\begin{table}
\caption{Interstellar reddening derived from DIBs, following
\citet{herbig75}.}
\label{dib}
\begin{center}
\begin{tabular}{lcccccccc}
\noalign{\smallskip}    \hline\noalign{\smallskip}
DIB             &$a$    &$b$    &EW(m\AA)       &$E(B-V)$       &N      \\
\noalign{\smallskip}    \hline\noalign{\smallskip}
6203 \AA        &280    &6      &410$\pm$95     &1.44$\pm$0.35  &28     \\
6269 \AA        &156    &70     &410$\pm$100    &2.18$\pm$0.65  &23     \\
6276-79 \AA     &189    &56     &325$\pm$75     &1.42$\pm$0.39  &26     \\
6613 \AA        &231    &43     &425$\pm$65     &1.65$\pm$0.28  &33     \\
\hline\noalign{\smallskip}
Mean            &       &       &               &1.67$\pm$0.35  &       \\
Weighted mean   &       &       &               &1.58$\pm$0.18  &       \\
\noalign{\smallskip}   \hline       
\end{tabular}
\end{center}
\end{table}

\subsection{Rotational velocity}

We estimated the projected rotational velocity, $v \sin i$, where $i$ is
the orbit inclination angle with respect to the line of sight, by measuring
the width of neutral Helium lines using the calibration by \cite{steele99}
and  by Fourier transform of line profiles \citep{simon07}.

For the first method, we used the WHT spectra taken on the night of 4 August
2015 as it provides the highest spectral resolution. Two spectra were
obtained on that night. We fitted a Gaussian profile and derive the FWHM of
four \ion{HeI}\ lines (4026 \AA, 4143 \AA, 4387 \AA,and 4471 \AA) on each
one of the two spectra as well as on the average. We repeated the procedure
three times, corresponding to different selections of the continuum. In
total we performed 36 measurements.  The final rotational velocity is the
mean of all these measurements.  The simple average and standard deviation
gave  $v \sin i=195\pm20$ km s$^{-1}$, whereas  the weighted average $v
\sin i=190\pm2$ km s$^{-1}$. 

\begin{figure}
\begin{center}
\includegraphics[width=8cm]{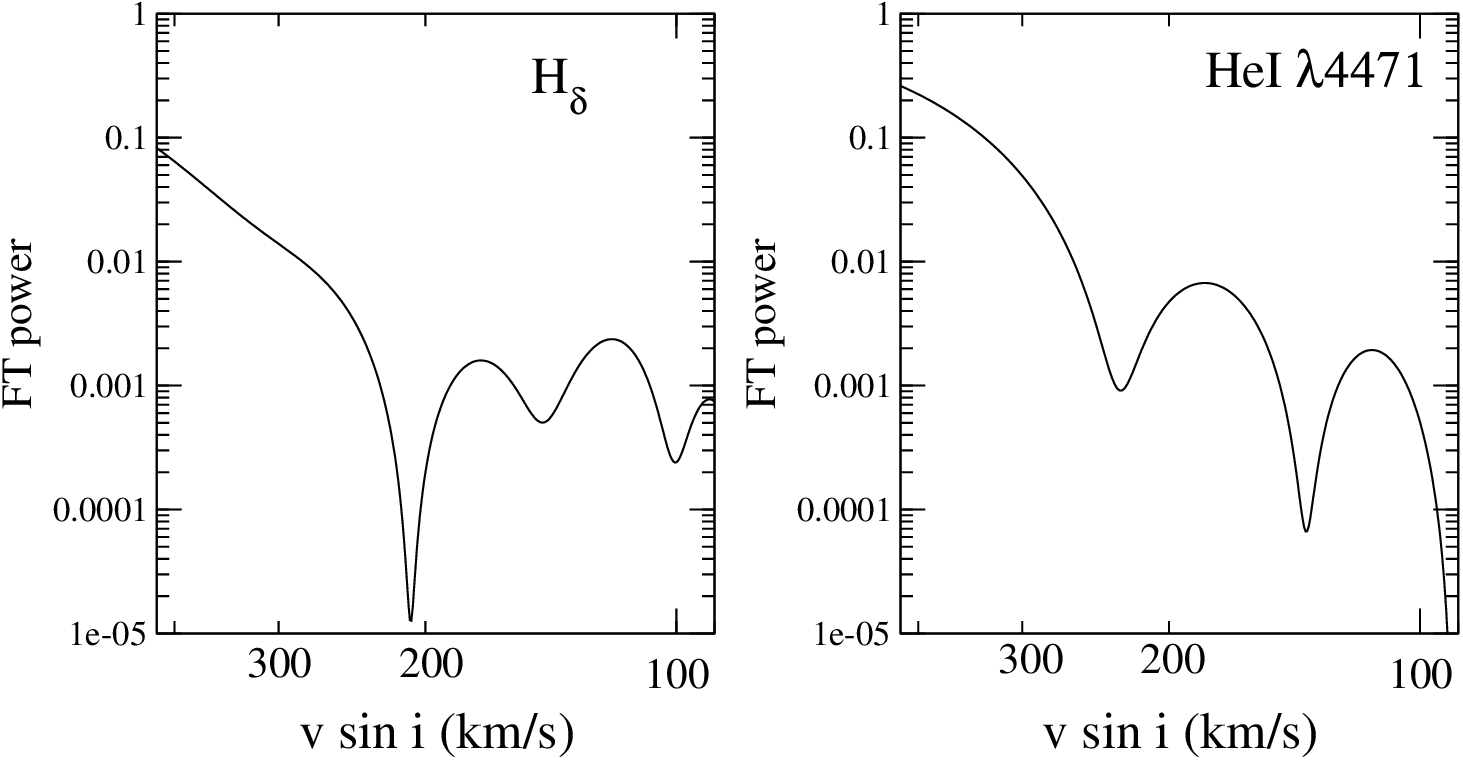}
\caption[]{Fourier transform of the profiles H$\delta$ and HeI at 4471 \AA\ lines. 
The first minimum gives the rotational velocity of the star. }
\label{ftvel}
\end{center}
\end{figure}

The rotational velocity can also be estimated by calculating the Fourier
transform of line profiles. The first zero in the frequency domain is
related to the rotational velocity of the star through the equation $v \sin
i=0.66\,c/(\lambda\,\sigma$), where $\lambda$ is the central wavelength of
the line and $\sigma$ the frequency at which the first zero is found
\citep{simon07}. This method works very well when the line profile is
dominated by rotation, but presents some limitations when other broadening
mechanism (macro- and micro-turbulence, Stark broadening) are at play. It
also requires high dispersion and high S/N spectra.  Figure~\ref{ftvel}
shows the Fourier transform of the WHT spectrum for two line profiles. The
average rotational velocity from hydrogen lines ($H\gamma$ and $H\delta$)
and helium lines (3926 \AA, 4026 \AA, 4387 \AA, and 4471 \AA) is $v \sin
i=207\pm12$ km s$^{-1}$ and $v \sin i=230\pm15$ km s$^{-1}$, respectively.
The reason for the higher values of the rotational velocity calculated from
the He I lines is that these lines are additionally broadened by the Stark
effect. 

Table~\ref{rotvel} gives the rotational velocity of some Be/X-ray binaries
obtained from the reference in column 8. Column 6 gives the inclination
angle of the orbit with respect to the line of sight, which allows us to
estimate the true rotation velocity of the Be star.  The term {\em shell}
refers to double-peak lines whose central depression  is lower than the
stellar continuum \citep[see e.g.][]{hummel00}. This type of profile occurs
in systems with high inclination angles. 

The rotational velocity of \sax\ is among the lowest measured in BeXB. This
low value, together with the fact that no shell profiles are observed in
the \ha\ line, suggest a relatively low inclination angle. On the other
hand, a double-peak profile is always seen, even when the equivalent width
is large. Thus, the inclination angle cannot be too low because we would
expect single-peak profiles, such as V\,0332+53
\citep{negueruela99,reig16}. A rough estimate of the inclination angle can
be obtained assuming that the Be star rotates close to critical velocity.
Be stars are fast rotators. They have, on average, larger observed
rotational velocities than B stars as a group \citep{slettebak82}. Even
though at present there is no consensus on how close to their critical
velocity Be stars rotate,  observations suggest that a large percentage of Be
stars rotate at 70--80\% of the critical value
\citep{slettebak82,porter96,yudin01}. The critical velocity depends on the
spectral type. For Be/X-ray binaries whose spectral type distribution spans
a very narrow range (O9-B2), the critical or break-up velocity is 500-600
km s$^{-1}$ \citep{townsend04,cranmer05}. Hence, if $v=(0.7-0.8)\times 550$
km s$^{-1}$, then $i\approx 25-35^{\circ}$.

\begin{table*}
\begin{center}
\caption{Comparison of \sax\ with other Be/X-ray binaries. }
\label{rotvel}
\begin{tabular}{lllccccc}
\noalign{\smallskip} \hline \noalign{\smallskip}
X-ray           &Optical        &Spectral&Disk-loss     &P$_{\rm orb}$  &Inclination    &$v \sin i$ &Reference \\
source          &counterpart    &type   &episodes       &(days)         &angle ($^{\circ}$)&(km s$^{-1}$)& \\
\noalign{\smallskip} \hline \noalign{\smallskip}
\sax\           &--             &B0V    &no     &262.6  &25--35         &195$\pm$20     &This work \\
4U 0115+634     &V635 Cas       &B0.2V  &yes    &24.3   &43             &300$\pm$50     &1         \\
RX J0146.9+6121 &LS I +61 235   &B1V    &no     &--     &--             &200$\pm$30     &2         \\
V 0332+53       &BQ Cam         &O8-9V  &no     &34.2   &$<$10          &$<$150         &3         \\
X Per           &HD 24534       &O9.5III&yes    &250    &23-30          &215$\pm$10     &4,5         \\
RX J0440.9+4431 &LS V +44 17    &B1III-V&yes    &150    &--             &235$\pm$15     &6,7    \\
1A 0535+262     &HD 245770      &O9.7III&yes    &111    &28-35          &225$\pm$10     &8,9         \\
IGR J06074+2205 &--             &B0.5IV &yes    &--     &--             &260$\pm$20     &10         \\
RX J0812.4-3114 &LS 992         &B0.5III-V&yes  &81.3   &--             &240$\pm$20     &11     \\
1A 1118-615     &Hen 3-640      &O9.5IV &no     &24     &15             &$\sim$300              &12,13         \\
4U 1145-619     &V801 Cen       &B0.2III&no     &187    &$<$45          &250$\pm$30     &14,15  \\
4U 1258-61      &V850 Cen       &B2V    &yes    &132    &shell$^\dag$   &$<$600         &16         \\
SAX J2103.5+4545&--             &B0V    &yes    &12.7   &--             &240$\pm$20     &17     \\
IGR\,J21343+4738&--             &B1IV   &yes    &--     &shell$^\dag$   &365$\pm$15     &18     \\
\noalign{\smallskip} \hline
\multicolumn{8}{l}{$\dag$: Shell stars are believed to be equator-on systems, i.e. $i\sim90^{\circ}$}\\
\multicolumn{8}{l}{[1] \citet{negueruela01a}, [2] \citet{reig97b}, [3] \citet{negueruela99}, [4] \citet{lyubimkov97}} \\
\multicolumn{8}{l}{[5] \citet{delgado01}, [6] \citet{reig05b}, [7] \citet{ferrigno13}, [8] \citet{haigh04} }   \\
\multicolumn{8}{l}{[9] \citet{grundstrom07b}, [10] \citet{reig10b}, [11] \citet{reig01}, [12] \citet{janot81}}    \\
\multicolumn{8}{l}{[13] \citet{staubert11}, [14] \citet{janot82}, [15] \citet{webster74}, [16] \citet{parkes80}}     \\
\multicolumn{8}{l}{[17] \citet{reig04}, [18] \citet{reig14a}}    \\
\end{tabular}
\end{center}
\end{table*}

\begin{figure}
\includegraphics[width=8cm]{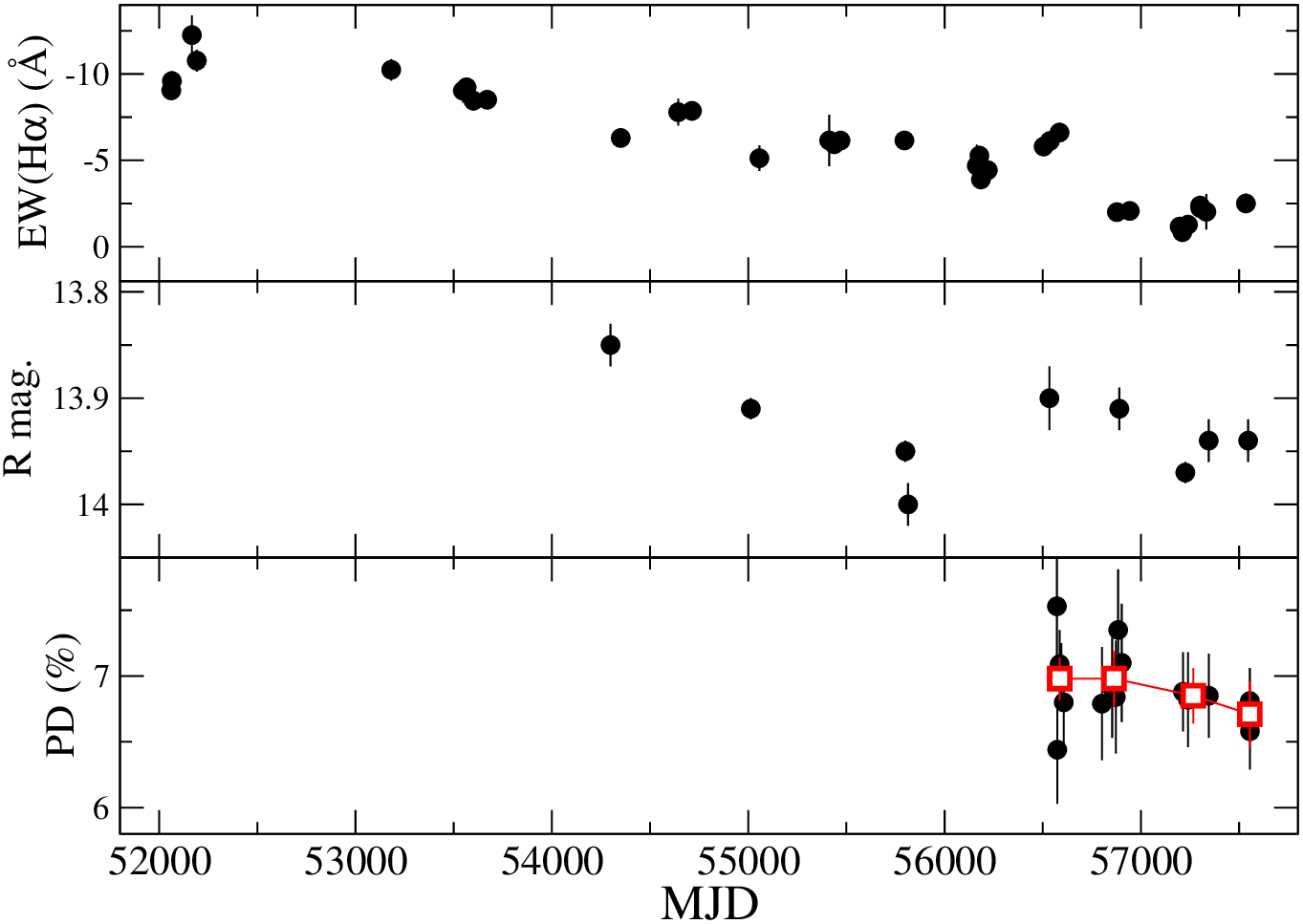} 
\caption[]{\ha\ equivalent width (top),  R-band magnitude (middle) and
R-band polarization degree (bottom) as a function of time. The squares in the
bottom panel represent the weighted average of the observations taken through 
one year from 2013 to 2016.}
\label{ewRP}
\end{figure}

\begin{figure}
\includegraphics[width=8cm]{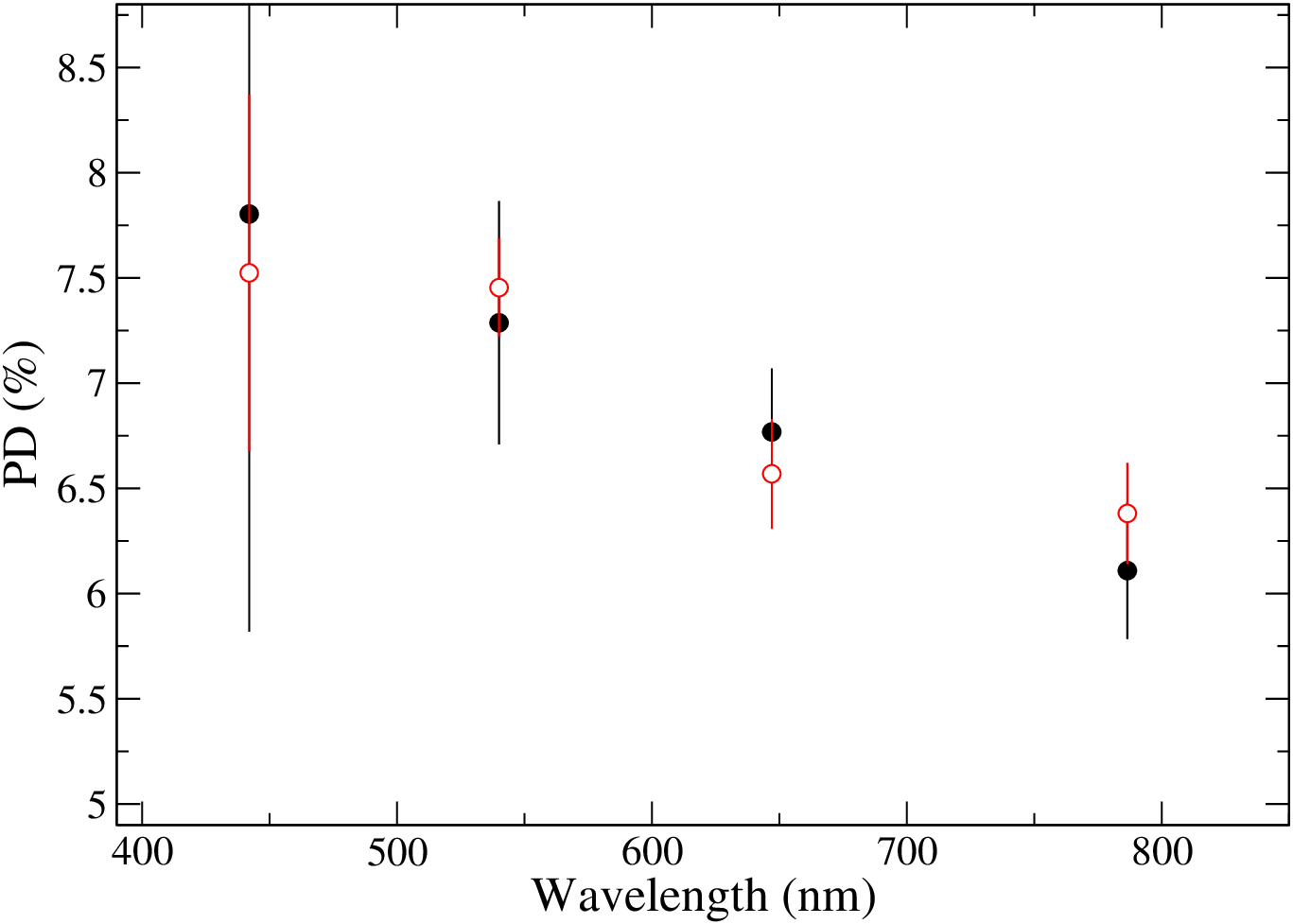} 
\caption[]{Dependence of polarization degree with wavelength. Black
(filled) and red (empty) points represent the 2015 and 2016 observations,
respectively (see Table~\ref{phot}).}
\label{polarimetry}
\end{figure}
\begin{figure}
\includegraphics[width=8cm]{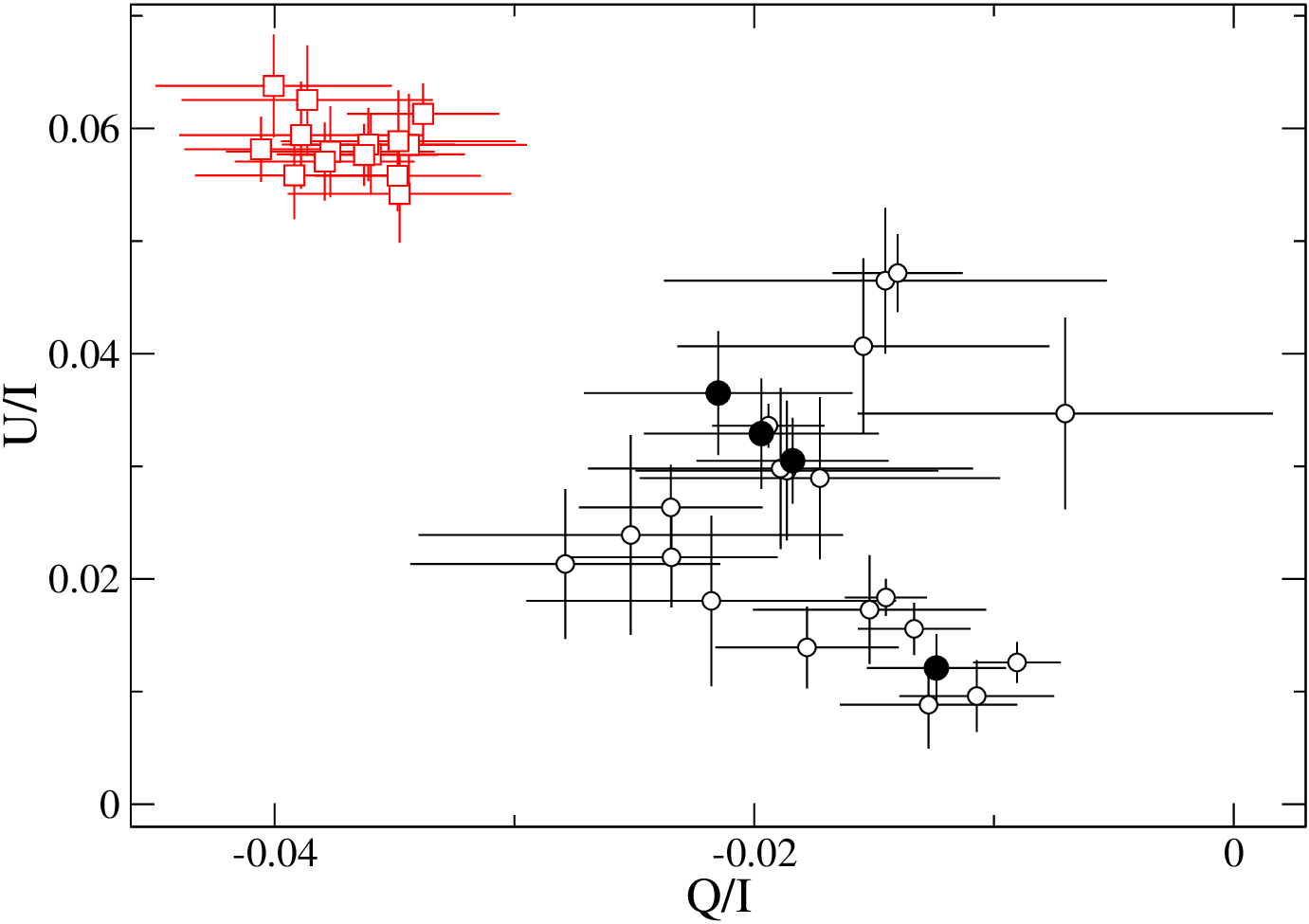} 
\caption[]{ R-band $q-u$ plane showing the field stars (circles) and \sax\
(squares). Filled circles correspond to field stars measurements  in the mask, 
while open circles are multi-epoch measurements outside of the mask.}
\label{qu-plot}
\end{figure}

\begin{table*}
\caption{Polarization details of four field stars in the vicinity of \sax\ 
in the $R$ band. $\rho$ is the angular distance from the source.}
\label{pol-fs}
\begin{center}
\begin{tabular}{lcccccccc}
\noalign{\smallskip}    \hline\noalign{\smallskip}
Date &  JD (2,400,000+) &RA     &DEC    &$\rho$($\arcmin$)              &PD(\%)         &PA ($^\circ$)      &$q$                    &$u$     \\
\noalign{\smallskip}    \hline\noalign{\smallskip}
23-05-14 &56801.575     &22h39m21.1s   &+61d16m40.3s &0.22      &3.6$\pm$0.4  &60.6$\pm$3.4  &-0.0184$\pm$0.0040     &0.0305$\pm$0.0038       \\
23-05-14 &56801.580     &22h39m14.9s   &+61d17m09.0s &1.0               &4.2$\pm$0.6  &60.3$\pm$3.8  &-0.0215$\pm$0.0056     &0.0365$\pm$0.0055       \\
19-11-15 &57346.298     &22h39m30.4s   &+61d15m46.6s &1.3       &3.8$\pm$0.5  &60.4$\pm$3.6  &-0.0197$\pm$0.0049     &0.0329$\pm$0.0049       \\
19-11-15 &57346.289     &22h39m12.6s   &+61d17m12.9s &1.2       &1.7$\pm$0.3  &67.9$\pm$4.9  &-0.0124$\pm$0.0029     &0.0121$\pm$0.0030       \\
\noalign{\smallskip}   \hline       
\end{tabular}
\end{center}
\end{table*}

\subsection{Polarization degree}

The continuum polarization in Be stars is attributed to Thomson scattering
of unpolarized starlight in the disk \citep{poeckert79}. The measured
polarization degree in \sax\ is well above the average observed in
classical Be stars. The maximum polarization level in an axisymmetric
circumstellar disk predicted by single-scattering plus attenuation models
is about 2\% \citep{waters92}. When multi-scattering is taken into account,
this level can increase to 3--4\% \citep{wood96,halonen13a}. In \sax, the
polarization degree is of the order of 7\% (Table~\ref{pol}). However, this
value includes the contribution of the interstellar medium (ISM).

To investigate the question of whether the measured polarization can be accounted
for entirely by the ISM or else some amount comes from the
source, we proceeded in various ways: {\em i)} perform multi-colour
polarimetry, {\em ii)} use average relationships between extinction and
polarization degree in the Galaxy, {\em iii)}  measure the polarization of
field and nearby stars around the X-ray source, and {\em iv)} perform a
variability analysis:

{\em i)} The wavelength dependence of interstellar polarization  is
observed to obey an empirical relation \citep{serkowski75},

\begin{equation}
P(\lambda)/P_{\rm max}=\exp[-k \ln^2(\lambda_{\rm max}/\lambda)]
\label{serk}
,\end{equation}

\noindent where $k$ determines the width or sharpness of the curve, $\lambda_{\rm
max}$ is the wavelength at which the polarization is maximum and is
directly related to the size of the dust grains \citep{coyne74,serkowski75}
and the total to selective extinction $R=A_V/E(B-V)$ \citep{whittet78}. The
mean value of $R=3.05$ corresponds to $\lambda_{\rm max}=0.545$ $\mu$m
\citep{whittet78}. Although Fig.~\ref{polarimetry} does not show a peak in
polarization at or around that wavelength, as expected if the variation of
the polarization degree with wavelength came exclusively from the
ISM, the large error bars and the small number of points
makes this result inconclusive. On the other hand, the polarization degree
of unreddened Be stars peaks in the blue ($\lambda \approx 0.45$ $\mu$m)
and decreases with wavelength in the range 0.45-0.80 $\mu$m
\citep{poeckert79,mcdavid01,halonen13a,haubois14}. Fig.~\ref{polarimetry}
agrees with the expected behaviour of Be stars.

{\em ii)} The relationship between polarization and extinction has been
studied by a number of authors
\citep{hiltner56,serkowski75,jones89,fosalba02}. These studies show that
the polarization fraction increases as the extinction increases, albeit
with a large scatter. Using the relationship $P_{\rm max, ISM}(\%)=3.5\,
E(B-V)^{0.8}$ \citep{fosalba02}, we estimate that the maximum contribution
of the ISM to the measured  optical polarization towards \sax\ would
be $P_{\rm ISM}=5.5$\%, which is lower than the observed polarization.

{\em iii)} An upper limit on the polarization degree from the ISM can be
estimated by measuring the polarization degree of field stars. If, on
average, the polarization degree of field stars is lower than that of the
source, then we can conclude that the ISM cannot account for all the
measured polarization, hence some of it must be intrinsic. We measured the
$q$ and $u$ Stokes parameters  in the R band of 20  stars in the
Robopol field of view not blocked by the mask in the target images and
derive the mean polarization degree ($p_{\rm fs}$) and angle ($\chi_{\rm
fs}$) and the corresponding standard deviations ($\sigma_p$ and
$\sigma_{\chi}$).  We selected those stars with more than five
measurements; the four light spots free of nearby sources to avoid
overlapping; signal-to-noise ratio, $p_{\rm fs}/\sigma_{p}$, larger than 3;
a dispersion in the polarization angle, $\sigma_{\chi}$, lower than
10$^{\circ}$; and a dispersion in polarization degree, $\sigma_p$, lower
than 1\% to avoid variable sources.  In addition, to increase the accuracy
of the polarization parameters, we selected four field stars of comparable
brightness in the vicinity of the target and observed them following the
same procedure as the target, that is, by placing them at the centre of the
mask of the polarimeter. Table~\ref{pol-fs} gives the details of these
observations. Figure~\ref{qu-plot} shows the $q-u$ plane of the field stars
(circles) and the target (squares). Filled circles correspond to the field
stars that were observed at the centre of the mask. We can estimate the
intrinsic polarization of \sax\ by vectorially subtracting the weighted
mean of the $q$ and $u$ Stokes parameters of the field stars
$(-0.0148\pm0.0008,0.0213\pm0.0007)$ from those of the source
$(-0.0368\pm0.0010,0.0583\pm0.0009)$. The polarization degree of the Be
star companion in \sax\ is $p=\sqrt{q^2+u^2}=4.3\pm0.1$\%.

{\em iv)} The polarization caused by the ISM is supposed to
be constant. Therefore, if variability is observed, then it can be
attributed to the source. Fig. \ref{ewRP} shows the variation of the \ha\
equivalent width, R magnitude, and polarization degree. All three quantities
decrease with time. A linear regression to all polarimetric data points
taking the errors into account does
not reveal significant variation ($p=0.2$, $\rho=-0.3$) and the slope is
almost consistent with zero ($s=-0.00026\pm0.00020$), i.e. no variation.
If the regression is performed on the weighted average, then the decrease
is significant at 95\% level ($p=0.05$, $\rho=-0.95$,
$s=-0.00027\pm0.00007$). Here $p$ is the probability that the correlation
occurs by chance and $\rho$ is the correlation coefficient.

In summary, although the results presented above are affected by large
uncertainty when they are considered individually, overall they provide
evidence that the light of the optical counterpart to \sax\ is polarized to
a few percent.

\begin{figure}
\includegraphics[width=8cm]{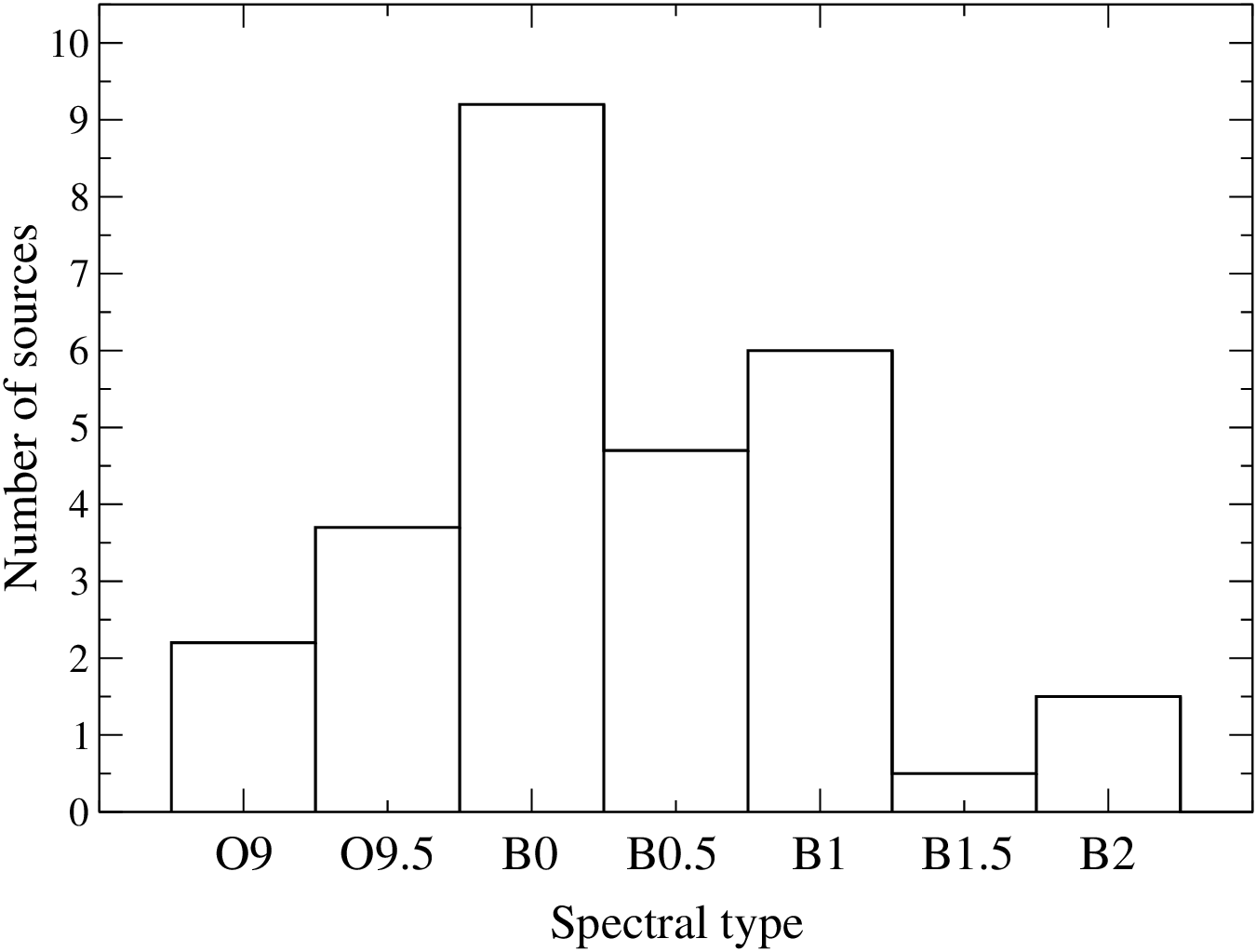} 
\caption[]{Distribution of spectral types of the optical counterparts in
BeXB. }
\label{spdist}
\end{figure}

\section{Discussion}

The information on \sax\ is scarce. Dedicated observations have been
reported on only three works: X-ray observations that led to the discovery
of \sax\ as a transient X-ray source and follow-up optical observations
\citep{intzand00}, discovery of X-ray pulsations and confirmation of the
orbital period from X-ray outbursts \citep{intzand01}, and one more optical
observation (photometry and spectroscopy) in the context of a global study
of circumstellar emission in BeXBs \citep{riquelme12}. In this work we
present the first detailed study of the optical variability of  \sax.  

The derived spectral type of B0V is in good agreement with the spectral
type distribution of confirmed galactic BeXBs, shown in Fig.~\ref{spdist}.
The data to create this figure were taken from  \citet{reig11}. When the
spectral type is only approximately known, we give the same weight to each
spectral subtype covered by the range. That is, if a star is classified as
B0-1V, then the bar corresponding to the spectral types B0, B0.5, and B1
each take one-third in the histogram of Fig.~\ref{spdist}. All spectrally
identified optical companions of confirmed BeXBs in the Milky Way have
spectral type earlier than B3 with a peak at B0  \citep[see
also][]{negueruela98b}.

The long-term decline of the H$\alpha$ equivalent width
(Fig.~\ref{hapar}) implies a progressive weakening of the disk over a
period of at least 15 years. The source appears to be slowly approaching a
disk-loss phase. A linear fit to the entire set of data in the top panel of
Fig.~\ref{hapar} yields that \ew\ decays at a relatively slow rate of
0.0015 \AA/d. Extrapolating this trend, we find that it would reach \ew=0
\AA\ on MJD 58400 (October 2018) and \ew=+2.5 \AA\ (the expected equivalent
width of a photospheric line of a B0V star) on MJD 60066 (May 2023). The
latter situation would correspond with the complete loss of the disk.  This
rate is typically 3--4 times slower than other BeXBs \citep{reig16} and can
be understood by the large orbital separation between the two components of
the binary.  \citet{reig16} found a correlation between the disk growth
rate and the orbital period. Systems with small orbits tend to display
faster growth rates and more optical variability. The interaction between
the disk and the neutron star appears as the primary cause of this
correlation.  \sax\ has the longest orbital period of all known BeXBs with
$\sim 262.6$ d \citep{intzand01}. In this system, the neutron star is so
distant that the Be star disk would be weakly affected by its
gravitational pull. In this sense, the disk in \sax\ would behave like those in classical isolated Be stars, which exhibit longer
dissipation-reformation cycles.

Although \ew\ has been increasing since 2015 (see
Table~\ref{spec}), which could be a sign that the disk has begun a full
recovery, we note the irregular evolution of the equivalent width. As can
be seen in Fig.~\ref{hapar} and Table~\ref{spec}, small increases in \ew,
for example in 2001, 2008, and 2013, were followed by deeper drops. Future
observations will reveal whether \sax\ will go through a full disk-loss
episode, or indeed the disk reached a minimum in 2015 and will start a new
reformation phase. The recovery of the disk without a complete loss  has
been seen in other BeXBs, such as GRO\,J2058+42 and RX J0146.9+6121
\citep[see e.g. Fig. 1 in][]{reig16}.

The symmetric profiles suggest the absence of a perturbed disk. This can be
explained assuming a relatively small disk. \citet{reig10b} have suggested that
a small disk  cannot support a density perturbation because the disk became
too tenuous. As a rough estimate, they found that during the final stages of
the disk evolution, V/R asymmetries can be seen up to a disk radius of
$\sim2 \, R_*$, whereas during the subsequent disk formation after a
disk-loss episode, a density perturbation does not develop until the disk
radius is of the order of 4 $R_*$. The disk radius can be estimated from
the separation of the emission line peaks and the rotational velocity
\citep{huang72},

\begin{equation}
\label{rad}
\frac{r_{\rm d}}{R_*}= \left(\frac{2 v_* \sin i}{\Delta V}\right)^2 
.\end{equation}

\noindent Using the values derived in the previous sections, the disk
radius in \sax\ varies in the range 1--6 $R_*$. The larger disks correspond
to observations prior 2005. From 2005-2009 the disk radius is $r_{\rm
d}\sim 2-3\, R_*$, while since 2010,  $r_{\rm d}\simless 2\, R_*$.

We also carried out a polarimetric study of \sax. We found that the optical
light emitted from the source is polarized and that the ISM
cannot account for the total measured polarization  (Fig.~\ref{qu-plot}).
We estimated the intrinsic polarization degree to be $\sim 4$\% in the R
band. Although in the limit of what models predict \citep{wood96}, this
value is consistent with electron scattering from the circumstellar disk.
If the origin of polarization is the circumstellar disk, then we would
expect long-term polarimetric variability, similar to the photometric and
spectroscopic variability. A correlation between polarization, \ha\
emission, and brightness has been reported for the Be/X-ray binary X Per and
interpreted in terms of disk development \citep{kunjaya95,roche97}. In
contrast, \citet{larionov87} suggested that the source of variable
polarized radiation measured in the BeXB 1A\,0535+262 is an accretion disk
around the neutron star. In \sax, the changes in polarimetry are less
pronounced than in spectroscopy and photometry, which show a large
amplitude decrease trend over time (Fig.~\ref{ewRP}). There are several
reasons for this. First, the time span covered by the polarimetric
observations is shorter and corresponds to recent years, where the
disk has presumably shrunk. Second, we argued above that \sax\ is viewed at
a small inclination angle. The polarization degree is strongly dependent on
the inclination angle. In pole-on systems, we do not expect to observe
polarization because of the uniform distribution of polarizing planes
\citep{wood96}. As the inclination angle increases, so does the
polarization degree, reaching a maximum at around 70--80$^\circ$
\citep{wood96,halonen13a}. Low-inclination systems ($i \simless
30^{\circ}$) typically exhibit  smaller amplitude changes in polarization
degree than higher inclination systems \citep{halonen13c,haubois14}.

\section{Conclusion}

We have performed optical photometric, spectroscopic, and polarimetric
observations of the optical counterpart to \sax. We derived a spectral type B0Ve from the ratios of various
metallic lines and we estimated the rotation velocity of the
underlying B star in 200 km s$^{-1}$ from the width of
hydrogen and He I lines.  We
estimated the distance to be $\sim$ 4.9 kpc from the photometric magnitudes and
colours and the strength of various diffuse interstellar bands. We report, for the first time,
intrinsic optical linear polarization from the circumstellar disk of the Be
star companion at a level of $4\%$. The long-term optical variability of
this system is characterized by the slow dissipation of the circumstellar
disk around the Be star companion. The observational consequence of this
decline is the decrease of the optical brightness, the strength of the \ha\
line, and the polarization degree.  We argue that the long variability timescales observed in this system are due to the relatively weak gravitational
pull exerted by the neutron as a consequence of its wide orbit. 

\begin{acknowledgements}

Skinakas Observatory is a collaborative project of the University of Crete
and the Foundation for Research and Technology-Hellas. The RoboPol project
is a collaboration between the University of Crete/FORTH in Greece, Caltech
in the USA, MPIfR in Germany, IUCAA in India, and Torun Centre for
Astronomy in Poland. The WHT and its service programme (service proposal
references SW2014b12) are operated on the island of La Palma by the Isaac
Newton Group in the Spanish Observatorio del Roque de los Muchachos of the
Instituto de Astrof\'{\i}sica de Canarias. This work has made use of NASA's
Astrophysics Data System Bibliographic Services and of the SIMBAD database,
operated at the CDS, Strasbourg, France.

\end{acknowledgements}

\bibliographystyle{aa}
\bibliography{../../artBex_bib}

\end{document}